\documentstyle[12pt,epsf]{article}
\def \thesection {\arabic{section}.}
\def \thesubsection {\thesection\arabic{subsection}.}

\def \sect #1 {\setcounter{equation} 0\section{#1}}
\def \appendix #1#2 {\par\bigskip\bigskip\noindent
                    {\Large {\bf Appendix {#1}. {#2} }}
                    \def\thesubsection{}
                    \def\theequation{{#1}.\arabic{equation}}
                    \setcounter{equation} 0 \par\bigskip\noindent}
\def \theequation {\arabic{equation}}             
\def \be  {\begin{equation}}
\def \ee  {\end{equation}}
\def \ba  {\begin{eqnarray}}
\def \ea  {\end{eqnarray}}
\def \baa {\begin{eqnarray*}}
\def \eaa {\end{eqnarray*}}
\def \bb  {\begin{thebibliography}}
\def \eb  {\end{thebibliography}}
\newcommand \ci [1] {\cite{#1}}
\newcommand \bi [1] {\bibitem{#1}}
\def \lab #1 {\label{#1}}
\newcommand\re[1]{(\ref{#1})}
\def \qqquad {\qquad\quad}

\newcommand\lr[1]{{\left({#1}\right)}}

\def \tr {\mbox{tr\,}}
\newcommand \vev [1] {\langle{#1}\rangle}

\newcommand \ket [1] {|{#1}\rangle}
\newcommand \bra [1] {\langle {#1}|}

\def \e {\mbox{e}}
\def \CO {{\cal O}}
\def \CP {{\cal P}}
\def \CT {{\cal T}}

\def\square{{\sqcup\kern-5.5pt\sqcap}}
\newcommand \widebar [1] {\overline{#1}}

\font\cmss=cmss10 \font\cmsss=cmss10 at 7pt
\def\inbar{\,\vrule height1.5ex width.4pt depth0pt}
\def\IC{\relax\hbox{$\inbar\kern-.3em{\rm C}$}}
\def\IZ{\relax\ifmmode\mathchoice
{\hbox{\cmss Z\kern-.4em Z}}{\hbox{\cmss Z\kern-.4em Z}}
{\lower.9pt\hbox{\cmsss Z\kern-.4em Z}}
{\lower1.2pt\hbox{\cmsss Z\kern-.4em Z}}\else{\cmss Z\kern-.4em Z}\fi}
\def\IR{{\hbox{{\rm I}\kern-.2em\hbox{\rm R}}}}
\def\R{{\tiny \IR}}
\def\IP{{\hbox{{\rm I}\kern-.2em\hbox{\rm P}}}}

\newcommand{\as}{\ifmmode\alpha_{\rm s}\else{$\alpha_{\rm s}$}\fi}
\def \alpi {\frac \as \pi}
\def \MS {\widebar{\rm MS}}

\newcommand\path{{\begin{figure}[h]
\vspace*{-19.5mm}
\hspace*{96mm}
\epsffile{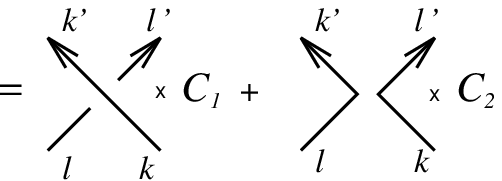}

\vspace*{-5mm}
\end{figure}}}

\begin{document}

\def\thefootnote{\fnsymbol{footnote}}
\thispagestyle{empty}
\hfill\parbox{50mm}{{\sc LPTHE--Orsay--96--53} \par
                         hep-ph/9607229  \par
                         June, 1996}
\vspace*{38mm}
\begin{center}
{\LARGE Evolution equation for gluon Regge trajectory}
\par\vspace*{15mm}\par
{\large I.~A.~Korchemskaya and  G.~P.~Korchemsky}%
\footnote{On leave from the Laboratory of Theoretical Physics,
          JINR, Dubna, Russia}
\par\bigskip\par\medskip

{\em Laboratoire de Physique Th\'eorique et Hautes \'Energies%
\footnote{Laboratoire associ\'e au Centre National de la Recherche
Scientifique (URA D0063)}
\\
Universit\'e de Paris XI, Centre d'Orsay, b\^at. 211\\
91405 Orsay C\'edex, France}

\end{center}
\vspace*{15mm}

\begin{abstract}
Analysing the asymptotic behaviour of the quark-quark elastic scattering
amplitude at high energy and fixed transferred momentum and assuming
that gluon is reggeized, we obtain the evolution equation for the gluon 
Regge trajectory in QCD. It is closely related to the renormalization 
properties of the Wilson lines and it is in a complete agreement with
the recent results of two-loop calculations.
\end{abstract}
\newpage
\def\thefootnote{\arabic{footnote}}
\setcounter{footnote} 0


\section{Introduction}

Understanding of the asymptotics of hadronic scattering amplitudes
at high energy $s$ and fixed transferred momentum $-t \ll s$
still remains an important challenge to perturbative QCD.
One of the main difficulties stems from the fact that in this
limit one should expect the emergence of a new regime of QCD \ci{Gr},
in which, in accordance with the Regge model expectations, partons
(quarks and gluons) as elementary excitations are replaced by new
collective degrees of freedom -- Regge poles, among which the two
families -- Reggeons and Pomerons with the quantum numbers of gluons 
and vacuum, respectively, should play a special role providing the 
leading asymptotic behaviour of the scattering amplitudes.

The previous attempts to solve the problem in the leading logarithmic
approximation (LLA), $\as \ll 1$ and $\as \ln s \sim 1$, have led
to the discovery of the BFKL formalism \ci{BFKL}. It was found that the gluon
is reggeized in the LLA and the QCD Pomeron appears in this
approximation as a compound state of two reggeized gluons. At high
energy and fixed transferred momentum, $-t \ll s$, the contribution
of the reggeized gluon, or {\it Reggeon\/},
to the scattering amplitude has a form $\sim
s^{\alpha_{_{\R}}(t)}$ with the Reggeon trajectory $\alpha_\R=\alpha_\R(t)$
given in the LLA by \ci{BFKL}
\be
\alpha_\R(t)-1=\omega(q_\perp^2)=-\as N_c
\lambda^{-2\varepsilon}
\int \frac{d^{2+2\varepsilon}k_\perp}{(2\pi)^{2+2\varepsilon}}
\frac{q_\perp^2}{k_\perp^2 (q-k)_\perp^2} + \CO(\as^2)\,,
\quad t=-q_\perp^2\,.
\lab{kt}
\ee
Since the Reggeon carries the color charge of gluon, its trajectory
is infrared (IR) divergent and one applies the dimensional regularization with
$D=4+2\varepsilon$ for $\varepsilon > 0$ and dimensional parameter
$\lambda^2$. At zero transferred momentum the Reggeon trajectory satisfies
$\alpha_\R(0)=1$ while for $t < 0$ one subtracts IR poles in
$\varepsilon$ in the $\MS$ scheme to find the renormalized expression
\be
\omega(q_\perp^2)=-\frac{\as}{2\pi}N_c \ln\frac{q_\perp^2}{\lambda^2}
+\CO(\as^2)\,.
\lab{1-loop}
\ee
The dependence on the IR cutoff $\lambda^2$
and IR poles in $\varepsilon$
are cancelled in the color-singlet compound states of
the Reggeons after one takes into account the interaction between
Reggeons \ci{BFKL}.

However, the LLA can not be satisfactory approximation to QCD in the Regge
limit due to the following reasons. First, the total cross-sections grow
in the LLA as powers of the energy $s$ violating the unitarity bound.
The nonleading corrections are responsible for the restoration of the
unitarity and at the same time they should define the region of energies
in which the LLA result will be meaningful. The problem of calculating
unitarization corrections is still open and it reveals
the intriguing connection of QCD in the Regge limit
with completely integrable two-dimensional systems.

Second, in the LLA the scattering amplitudes get equally important
contributions from the region of large and small
gluon transverse momenta. In order to be able to control nonperturbative
effects associated with soft gluons one has to separate the latter
contribution from the scattering amplitudes by performing the
factorization procedure \ci{CSS}. In the LLA, this can be done in \re{kt}
by introducing the factorization scale $\mu$ such that
$q_\perp^2 \gg \mu^2 \gg \lambda^2$ and dividing the whole
interval of gluon transverse momenta into the regions of  ``soft''
momenta, $k_\perp^2 < \mu^2$, and ``hard'' momenta, $k_\perp^2 > \mu^2$.
These two regions provide the $\mu-$dependent additive contributions to
the Reggeon trajectory and the $\mu-$dependence cancels in their sum.
Beyond the LLA, one has to follow the general procedure \ci{CSS,BS89} in
order to obtain the factorized expression for the scattering
amplitude, in which the contribution of soft gluons
is described by a nonperturbative soft function $S$, while that of
hard gluons is absorbed into perturbatively calculable hard function $H$. In an
analogy with deep inelastic scattering, one may try to interpret the
factorization procedure as some kind of the Regge operator product
expansion (OPE), in which the soft function is associated with the
matrix element of
local composite operators and the hard function is defined as a
coefficient function. Having identified the relevant operators, one
should try to relate their anomalous dimensions with the
properties of the Regge trajectories in QCD \ci{K94,SS94}.

In this paper we propose the OPE for the Reggeon trajectory and
identify the relevant operators as Wilson lines.
We establish the evolution equation for the Reggeon trajectory and
show its close relation to the renormalization properties of the
so-called cross singularities of Wilson lines. Our approach is closely
related to the techniques developed by Lipatov \ci{L} and by Botts and
Sterman \ci{BS89}.

Recently, the next-to-leading corrections to the BFKL formalism became
available \ci{FL,FFK95,FFK96}. In particular, the 2-loop correction
to the Reggeon trajectory was calculated \ci{FFK95} and the result can be
represented in the following form \ci{FFK96}
$$
\omega(t)=\omega^{(1)}(t)+\omega^{(2)}(t) + \CO(\as^3)
$$
with the 1-loop correction $\omega^{(1)}$ defined in \re{1-loop} and
\be
\omega^{(2)}(t)=\lr{\alpi}^2 N_c \lr{w_1 L^2 + w_2 L + w_3}\,.
\lab{2-loop}
\ee
Here, $\as=\as(\lambda^2)$, $L=\frac12\ln(q^2/\lambda^2)$ and the
coefficients $w_i$ are given by
$$
w_1=\frac{11}{12}N_c - \frac16 n_f\,,
\quad
w_2=-\lr{\frac{67}{36}-\frac{\pi^2}{12}} N_c + \frac5{18} n_f\,,
\quad
w_3=\lr{\frac{101}{108}-\frac18 \zeta(3)} N_c - \frac7{54} n_f\,,
$$
where $n_f$ is the number of light quark flavours.
It was observed \ci{FFK96} that, as a result of nontrivial cancellations 
between
different contributions, the $L^3$ term vanishes in \re{2-loop} and the
coefficient in front of the $L^2$ term turns out to be proportional to the
one-loop $\beta-$function. As we will show in Sect.~4, the
evolution equation provides a simple understanding of these properties
as well as interpretation of the remaining coefficients.

\section{Operator production expansion}

The Reggeon trajectory should not depend on the type of the scattered
particles. Using this property one
can simplify the analysis by considering the elastic scattering
of two quarks with mass $m$ in the Regge limit of large
energy $s=(p_1+p_2)^2$ and fixed transferred momentum $t=(p_1-p_1')^2$.
Here, $p_1$, $p_2$ and $p_1'$, $p_2'$ are on-shell momenta of incoming
and outgoing quarks with the color indices $(i,j)$ and
$(i',j')$, respectively. We will consider the asymptotic behaviour
of the quark-quark scattering amplitude $T^{i'j'}_{ij}$ in the
following Regge kinematics
\be
s \gg m^2 \gg -t \gg \lambda^2
\lab{kin}
\ee
with the IR regulator chosen to be much larger the QCD scale,
$\lambda^2 \gg \Lambda_{\sc qcd}^2$. In this limit, the scattering
amplitude becomes a function of two ratios only \ci{K94}, $s/m^2$ and
$t/\lambda^2$, and, as a tensor in the color space, it can be
decomposed over the scalar components with the quantum numbers of
vacuum, $T^{(0)}$, and gluon, $T^{(8)}$, in two equivalent ways
\be
T^{i'j'}_{ij}
=T^{i'j'}_{ij}\lr{\frac{s}{m^2},\frac{t}{\lambda^2}}
=\delta_{i'i}\delta_{j'j}\ T^{(0)} + t^a_{i'i}\,t^a_{j'j}\ T^{(8)}
=\delta_{i'i}\delta_{j'j}\ T_1 + \delta_{i'j}\delta_{j'i}\ T_2\,,
\lab{flow}
\ee
where $t^a$ is the generator of the $SU(N_c)$ in the fundamental
representation and the following relations hold
$T^{(8)}= 2 T_2$ and $T^{(0)}=T_1+\frac1{N_c}T_2$.
If one assumes that the gluon is reggeized, then its contribution
to the negative signature
component $T_2^{-}$ of $T_2$ (or equivalently $T^{(8)}$) has the
following factorized form in the Regge
kinematics \re{kin}
\be
T_2^-\equiv
T_2\lr{\frac{s}{m^2},\frac{t}{\lambda^2}}
-T_2\lr{-\frac{s}{m^2},\frac{t}{\lambda^2}}
= \frac{\beta_\R^2(t)}{t}
  \left[
  \lr{\frac{s}{m^2}}^{\alpha_{_\R}(t)}-\lr{\frac{-s}{m^2}}^{\alpha_{_\R}(t)}
  \right]\,,
\lab{T2-}
\ee
where the residue factor $\beta_\R$ measures the coupling of the Reggeon
to the quark.

The incoming quarks scatter each other
by exchanging gluons in the $t-$channel with the total momentum
$q\approx q_\perp$ and $q_\perp^2=-t$. In the limit \re{kin}, quarks
behave effectively as heavy particles and the only effect of their
interaction with gluons is the appearance of the eikonal phase in the
heavy quark wave function. This phase is given by the Wilson
line, $\CP\exp(i\int_C dx \cdot A(x))$, evaluated along the classical
trajectory $C$ of the quark. Combining eikonal phases
of the incoming
quarks one can get the following representation for the scattering
amplitude \ci{VV,K94}
\be
T_{ij}^{i'j'} \lr{\frac{s}{m^2},\frac{q^2}{\lambda^2}}
=i\frac{s}{m^2} \int d^2z\, \e^{-i\vec z\cdot\vec q_\perp}
\langle 0|\CT\, W_+^{i'i}(0) W_-^{j'j}(z)|0\rangle\, ,
\qquad t=-{\vec q_\perp^{\,2}}\,,
\lab{T}
\ee
where $W_+$ and $W_-$ are the eikonal phases corresponding to the incoming
quarks with momenta $p_1$ and $p_2$, respectively:
$$
W_+(0)=\CP\exp\left(
       i\int_{-\infty}^{\infty}d\alpha\ v_1\cdot A(v_1\alpha)\right)\,,
\qquad
W_-(z)=\CP\exp\left(
       i\int_{-\infty}^{\infty}d\beta\ v_2\cdot A(v_2\beta+z)\right)\,.
$$
Here, integration is performed along two infinite lines in the direction of
the heavy quark velocities $v_i=p_i/m$ separated by the impact vector
$z=(0^+,0^-,z_\perp)$ in the transverse plane. In the longitudinal plane
defined by the quark momenta $p_1$ and $p_2$ one introduces the angle $\gamma$
between the quark trajectories as $\cosh\gamma = (v_1\cdot v_2)=
(p_1\cdot p_2)/m^2$ and in the limit \re{kin} one gets
$$
\gamma = \ln \frac{s}{m^2} \gg 1\,.
$$
According to \re{T}, the nontrivial $s-$dependence of the scattering amplitude
comes from the analogous dependence of the vacuum expectation value
of the product of two Wilson lines%
\footnote{In this relation one can replace the vacuum by a
hadronic state since the perturbative corrections to the matrix element
of the Wilson line operators are not sensitive to the particular choice
of nonperturbative state}
\be
W^{i'j'}_{ij}=W^{i'j'}_{ij}(\gamma,\lambda^2 z^2)=
\bra{0} \CT W_+^{i'i}(0) W_-^{j'j}(z)\ket{0}\,,
\lab{W}
\ee
which in turn is translated into the kinematical dependence of the
integration contour on the angle $\gamma$. The one-loop correction to
the Wilson line can be easily calculated using the dimensional
regularization as \ci{K94}
\be
W^{i'j'}_{ij}=\delta_{i'i}\delta_{j'j}+t^a_{i'i}\,t^a_{j'j}\,
\alpi (-i\pi\coth\gamma)\Gamma(\varepsilon)
(\pi\lambda^2 z_\perp^2)^{-\varepsilon}+\CO(\as^2)\,,
\lab{W-1}
\ee
where the pole in $1/\varepsilon$ has an IR origin.
Substituting this expression into \re{T} and performing the Fourier
transformation one finds that the IR pole disappears and one reproduces
the quark-quark scattering amplitude in the Born approximation
\be
\lr{T_{\rm Born}}^{i'j'}_{ij}
= \frac{g^2}{q_\perp^2} \frac{s}{m^2} \,
t^a_{i'i}\,t^a_{j'j}\,.
\lab{Born}
\ee
Calculating the quark-quark scattering amplitude to higher orders in
$\as$ one might expect that the Feynman integrals over gluon momenta 
should involve four momentum scales: $s^2$, $m^2$, $t$ and $\lambda^2$. 
In reality, due to the properties of the Wilson lines, the first two 
scales enter only via the $\gamma-$dependence and the relevant scales become
$\lambda^2$ and $1/z^2$. The first scale provides the IR cutoff for
gluon momenta and the second one, $1/z^2$, cuts the gluon momenta from 
above and has a meaning of an UV cutoff \ci{KR}.
Let us introduce the factorization scale $\mu$ such that $1/z^2 \gg
\mu^2 \gg \lambda^2$ and separate gluon momenta into hard region and
soft region. Then, an arbitrary Feynman diagram contributing to the
Wilson line takes a form shown on fig.~1, where according to the value
of their momenta gluons belong either to hard (H) or soft (S)
subprocesses. The contribution of the hard subprocess depends on the
factorization scale $\mu$ and on the large scale $1/z^2$ while the
contribution of the soft subprocess depends on $\mu$ and IR cutoff
$\lambda^2$. Both subprocesses depend also on the angle $\gamma$ as
well as on the color indices of the scattered quarks. The resulting
factorized expression for the Wilson line has the following form%
\footnote{In what follows the dependence of the subprocesses
on the coupling constant $\as(\mu^2)$ is implied}
\ci{BS89}
\be
W^{i'j'}_{ij}(\gamma,\lambda^2 z^2)
=S^{i'j';kl}_{ij;k'l'}(\gamma,\mu^2/\lambda^2)\
H^{k'l'}_{kl}(\gamma,\mu^2 z^2)\,,
\lab{W=HS}
\ee
where summation is performed over repeated color indices.
The color flow inside the hard subprocess can be decomposed similar
to \re{flow} as
\vspace*{2.5mm}
\be
H^{k'l'}_{kl}= \delta_{k'k}\delta_{l'l}\ C_1(\gamma,\mu^2 z^2)
             + \delta_{k'l}\delta_{l'k}\ C_2(\gamma,\mu^2 z^2)
\hspace*{60mm}
\lab{H}
\ee

\path

\noindent
where $C_1$ and $C_2$ are invariant coefficient functions.
The soft subprocess $S$ describes the coupling of the soft gluons to the
Wilson lines corresponding to the scattered heavy quarks. Inside the
hard subprocess the Wilson lines are separated
in the transverse plane at the short distance $z^2_\perp \ll 1/\mu^2$.
However, since the transverse momenta of gluons inside S is small,
$k_\perp^2 < \mu^2$, the soft gluons can not resolve short distances
inside H and calculating their contribution one can shrink the hard
subprocess into a point at $z=0$. Notice that at $z=0$ the
quark trajectories cross each other at the origin and two different
color flows inside the hard subprocess \re{H} lead to two possible
configurations of the Wilson lines describing the soft function $S$
\ba
\lr{W_1}^{i'j'}_{ij} &=& S^{i'j';kl}_{ij;kl}
               = \vev{0|\CT
                 \left[\CP \e^{i\int_\nwarrow 
                                       dx \cdot A(x)}\right]^{i'}_i
                 \left[\CP \e^{i\int_\nearrow 
                                       dx \cdot A(x)}\right]^{j'}_j
                           |0}\,,
\nonumber
\\
\lr{W_2}^{i'j'}_{ij} &=&  S^{i'j';kl}_{ij;lk}
               =  \vev{0|\CT
                 \left[\CP \e^{i\int_> dx \cdot A(x)}\right]^{i'}_j
                 \left[\CP \e^{i\int_< dx \cdot A(x)}\right]^{j'}_i
                           |0}\,.
\lab{W12}
\ea
The integration contours entering into these expressions are different
only at the vicinity of the cross point and they are shown by
solid lines on fig.~1.
\begin{figure}[ht]
\centerline{\epsffile{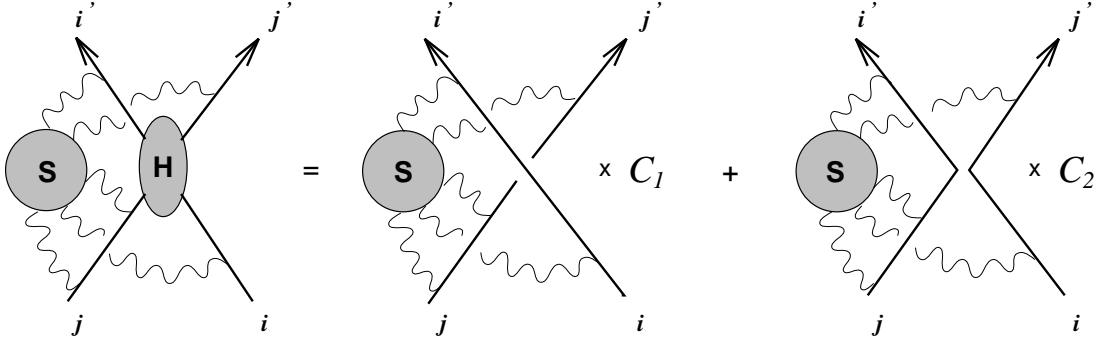}}
\caption{
The general structure of the Feynman diagrams for the scattering
of two Wilson lines at short transverse distances $z_\perp$.
The soft gluons are factorized into two Wilson lines with a 
cross point at the origin. The coefficient functions $C_1$ 
and $C_2$ represent the hard gluons contribution.}
\end{figure}

Finally, substituting \re{H} and \re{W12} into \re{W=HS}
we obtain the expression for
the Wilson line \re{W} in the form of the operator product expansion
\be
W^{i'j'}_{ij}(\gamma,\lambda^2 z^2) = C_1(\gamma,z^2 \mu^2)
\lr{W_1}^{i'j'}_{ij}(\gamma,\mu^2/\lambda^2)
+ C_2(\gamma,z^2 \mu^2)
\lr{W_2}^{i'j'}_{ij}(\gamma,\mu^2/\lambda^2) \equiv C \cdot W \,,
\lab{OPE}
\ee
where the notation was introduced for the doublets
\be
C=(C_1,\ C_2)\,, \qquad W= \lr{{W_1 \atop W_2}}\,.
\lab{doub}
\ee
To one-loop order, the calculation of the coefficient functions and
the Wilson lines in the $\MS$ scheme gives
\be
(C_1-1)N_c=-C_2=-\alpi \lr{i\pi\coth\gamma} 
\ln\lr{\mu|z_\perp|\e^{-\gamma_{\sc e}}/2}
+\CO(\as^2)
\lab{C}
\ee
with $\gamma_{\sc e}$ being the Euler constant and
\baa
W_1=\delta_{i'i}\delta_{j'j}
-\alpi\ln\frac{\mu}{\lambda}
&& \hspace*{-6mm}
\lr{i\pi\coth\gamma}\left[-\frac1{N_c}
\delta_{i'i}\delta_{j'j}+\delta_{i'j}\delta_{j'i}
\right] +\CO(\as^2)
\,,
\\
W_2=\delta_{i'j}\delta_{j'i}
-\alpi\ln\frac{\mu}{\lambda}
&& \hspace*{-6mm}
\bigg[
\hspace*{3mm}
\delta_{i'i}\delta_{j'j}\lr{\gamma\coth\gamma-i\pi\coth\gamma-1}
\\
&& \hspace*{-6mm}
+\left.
\delta_{i'j}\delta_{j'i}\lr{-N_c(\gamma\coth\gamma-1)
+\frac{i\pi}{N_c}\coth\gamma}
\right] +\CO(\as^2)\,.
\eaa
Substituting these expressions into \re{OPE} one reproduces
the one-loop result \re{W-1} for $W^{i'j'}_{ij}$ renormalized
in the $\MS$ scheme.

We stress that the Wilson line operator in the l.h.s.\ of
\re{OPE} is a nonlocal operator of the gauge fields both in the
longitudinal and transverse planes. The relation \re{OPE} states
that it can be expanded over two Wilson line operators which
are still nonlocal in longitudinal plane but local in the transverse
direction. We notice, however, that nonlocality of $W_1$ and $W_2$
in the longitudinal direction can be effectively eliminated if
one uses the representation for the Wilson lines as correlators of
local heavy quark fields in the effective heavy quark field theory \ci{KR}.

\section{Cross singularities of Wilson lines}

The scattering amplitude \re{T} and the Wilson line \re{OPE}
do not depend on an arbitrary
factorization scale $\mu$ while the coefficient functions $C$ and
the Wilson lines $W_{1,2}$ depend separately on $\mu$. In
a full analogy with the standard OPE, the latter dependence can be
found using the renormalization properties of the Wilson lines
$W_1$ and $W_2$ defined in \re{W12} and entering into the OPE
expansion \re{OPE}. These properties
have been studied in details in past \ci{ren} without
any relation to the high-energy scattering and they were found
to depend on the choice of the integration contours $C$ and on the
representation of the $SU(N_c)$ group in which the Wilson lines
are defined. The peculiar features of the Wilson lines $W_1$ and
$W_2$ are that, first, they are defined in the fundamental (quark)
representation and, second, the integration contours entering into
their definitions cross each other at the origin. As a result, the
Wilson lines $W_1$ and $W_2$ are mixed with each other under
renormalization and their $\mu-$dependence is described by the
following RG equation \ci{ren}
\be
\lr{\mu \frac{\partial}{\partial \mu} + \beta(g)\frac{\partial}{\partial
g} +\Gamma_{\rm cross}(\gamma,\as)} W(\gamma,\mu^2/\lambda^2)=0
\lab{RG}
\ee
with the doublet $W$ defined in \re{doub}. The evolution equation for the
coefficient functions can be easily obtained from \re{OPE} and \re{RG}
using $\mu-$independence of the product $C \cdot W$.
The equation \re{RG} introduces
into consideration a new object, a $2\times 2$ matrix of anomalous
dimensions $\Gamma_{\rm cross}(\gamma,\as)$. As we will show below,
this matrix governs the Regge behaviour of the quark-quark scattering
 amplitude in much the same way as the anomalous dimensions of local
twist 2 operators define the evolution of the structure function
of DIS.

The matrix of the cross anomalous dimensions $\Gamma_{\rm cross}(\gamma,\as)$
has the following interesting properties. It is gauge invariant and it
depends only on the coupling constant and the cross angle
$\gamma=\ln\frac{s}{m^2}$ but not on the color indices of the Wilson
lines. The matrix $\Gamma_{\rm cross}(\gamma,\as)$ was calculated
to two-loop order in QCD and its explicit expression as well as
some general properties can be found in \ci{KK95}. In application to
high-energy scattering, it is of most importance to know
the asymptotics of the matrix in the limit $\gamma=\ln\frac{s}{m^2}
\gg 1$. In this limit, to all order in $\as$, the different elements
of the matrix have the following behaviour 
\be
\Gamma_{\rm cross}^{11} \sim \Gamma_{\rm cross}^{12} \sim \gamma^0\,,
\qquad
\Gamma_{\rm cross}^{21} \sim \Gamma_{\rm cross}^{22} \sim \gamma\,,
\lab{asym}
\ee
which would imply, in general, the asymptotics $\det \Gamma_{\rm cross} \sim
\gamma$. However, the matrix elements have an interesting ``fine
structure'', due to which remarkable cancellation of the leading $\sim
\gamma$ term occurs leading to $\det \Gamma_{\rm cross} \sim
\gamma^0$ with $\tr \Gamma_{\rm cross} \sim \gamma$. As a result,
the eigenvalues $\Gamma_\pm$ of the matrix $\Gamma_{\rm cross}$
can be calculated as
\be
\Gamma_+ = \gamma\ \Gamma_{\rm cusp}(\as) + \CO(\gamma^0)\,,
\qquad
\Gamma_- = \CO(\gamma^{-1})\,,
\lab{+-}
\ee
with $\Gamma_+/\Gamma_-=\CO(\gamma^2)$.
Here, the leading coefficient $\Gamma_{\rm cusp}(\as)$ is closely
related to the so-called ``cusp'' anomalous dimension and it is
known to two-loop order as%
\footnote{In comparison with \ci{KK95}, we added here the two-loop
contribution of light quark flavours to $\Gamma_{\rm cusp}(\as)$. 
Recently, the special class of corrections to $\Gamma_{\rm cusp}(\as)$
due to quark loops, $\as(\as n_f)^k$, was resummed to all orders in 
$\as$ \ci{BB}}
\be
\Gamma_{\rm cusp}(\as)=\alpi N_c + \lr{\alpi}^2 N_c
\left[
N_c\lr{\frac{67}{36}-\frac{\pi^2}{12}}
-n_f\frac5{18}
\right]
+\CO(\as^3)
\lab{cusp}
\ee
where $n_f$ is the number of light quark flavours.

The RG equation \re{RG} was checked in \ci{KK95} by performing the
calculation of $W_1$ and $W_2$ to the order $\as^2$. It was observed
that the perturbative corrections to the Wilson lines vanish at $\mu=\lambda$
$$
W_1(\gamma,1)=\delta_{i'i}\delta_{j'j}\,,\qquad
W_2(\gamma,1)=\delta_{i'j}\delta_{j'i}\,.
$$
Using this relation as a boundary condition, we solve the RG
equation \re{RG}, obtain the expressions for the Wilson lines
and substitute them into \re{OPE} to get
\be
W^{i'j'}_{ij}(\gamma,\lambda^2 z^2) =
\lr{C_1(\gamma,\mu^2 z^2),C_1(\gamma,\mu^2 z^2)}\ \cdot
A(\gamma,\mu^2/\lambda^2) \ \cdot
\lr{{\delta_{i'i}\delta_{j'j} \atop \delta_{i'j}\delta_{j'i}}}\,,
\lab{sol}
\ee
where the notation was introduced for $2\times 2$ matrix
\be
A(\gamma,\mu^2/\lambda^2)= \CT \exp\lr{
-\int_\lambda^\mu \frac {d\tau}{\tau}
\Gamma_{\rm cross}(\gamma,\as(\tau^2))}\,.
\lab{A}
\ee
Here, the matrices $\Gamma_{\rm cross}(\gamma,\as(\tau^2))$ do not commute
for different values of $\tau$ and $\CT$ orders them along the integration
path. Except of the special cases \ci{SS94,K94}
(frozen coupling constant, one-loop
order approximation for $\Gamma_{\rm cross}$), the evaluation of the
matrix $A$ becomes extremely nontrivial.

\section{Evolution equation}

Substituting \re{sol} into \re{T} one performs the Fourier transformation
of the coefficient functions
\be
\widetilde C_{1,\,2}(\gamma,q_\perp^2/\mu^2) = 
i\frac{q_\perp^2}{4\pi} \int d^2 z_\perp\ \e^{i q_\perp\cdot z_\perp}
C_{1,\,2}(\gamma,z_\perp^2\mu^2)
\lab{Four}
\ee
and finally obtains the  representation for the scattering amplitude in the
form \re{flow} with
\be
(T_1(\gamma,q^2/\lambda^2), T_2(\gamma,q^2/\lambda^2))
=\sigma_0\
(\widetilde C_1(\gamma,q^2/\mu^2), \widetilde C_2(\gamma,q^2/\mu^2))
            \, \cdot A(\gamma,\mu^2/\lambda^2)\,,
\lab{T0}
\ee
where $\sigma_0=\frac{2\pi}{q_\perp^2}\frac{s}{m^2}$ 
and the matrix $A$ was defined in
\re{A}. In this relation, the
dependence of the invariant scattering amplitudes $T_1$ and $T_2$
on the IR cutoff and on the transferred momentum $q^2$ is factorized
into the matrix elements of the Wilson lines, $A$, and the coefficient
functions $\widetilde C$, respectively. In the matrix notations,
the doublet $\widetilde C$ and the matrix $A$ are renormalized
multiplicatively and they satisfy the RG equations involving the anomalous
dimension $\Gamma_{\rm cross}$. To one-loop order $\widetilde C_{1,2}$
can be obtained from \re{Four} using \re{C} as
\be
\widetilde C_1=-\frac1{N_c}\as +\CO(\as^2) \,,
\qquad
\widetilde C_2=\as +\CO(\as^2) \,.
\lab{C2t}
\ee

As a function of the IR cutoff, the matrix \re{A} satisfies the relation
$\lambda\frac{d}{d\lambda} A = A \Gamma_{\rm cross}$. Then,
differentiating the both sides of \re{T0} with respect to $\lambda$ we
obtain the evolution equation for the scattering amplitudes as
\be
\lambda\frac{d}{d \lambda}\
(T_1, T_2)
=
(T_1, T_2)\, \cdot
\Gamma_{\rm cross}(\gamma,\as(\lambda^2))\,.
\lab{evol}
\ee
This equation is a natural generalization of an analogous evolution
equation for the Sudakov form factor in QCD \ci{KR}. In the latter case,
all effects of soft gluon radiation are factorized into a single
Wilson line and the dependence of the Sudakov form factor on the IR cutoff
becomes in one-to-one correspondence with renormalization of
the ``cusp'' singularities of the Wilson line. As a result, the Sudakov
form factor satisfies the evolution equation similar to \re{evol} with
the matrix $\Gamma_{\rm cross}$ replaced by the ``scalar'' cusp anomalous
dimension $\Gamma_{\rm cusp}$.

To identify the contribution of the Reggeon to the scattering amplitude
\re{T2-} we exclude $T_1$ from the system \re{evol} and obtain the evolution
equation for $T_2$ as
\be
T_2'' + T_2'\ f + T_2\ \varphi=0
\lab{eq1}
\ee
with the coefficients $f$ and $\varphi$ defined as
\be
f=-\tr \Gamma_{\rm cross} - \lr{\ln\Gamma_{\rm cross}^{12}}'\,,
\qquad
\varphi= \det \Gamma_{\rm cross} - \lr{\Gamma_{\rm cross}^{22}}'
+\Gamma_{\rm cross}^{22} \lr{\ln\Gamma_{\rm cross}^{12}}'\,.
\lab{f}
\ee
Here, the prime denotes the derivative $\frac{d}{d\ln \lambda}$ and
the elements of $\Gamma_{\rm cross}(\gamma,\as)$ depend on $\lambda$
only through $\as(\lambda^2)$. We notice that the dependence on the energy $s$
enters into this equation via the matrix elements of the cross anomalous
dimension. Using the asymptotics \re{asym} and \re{+-}
we obtain the behavior of $f$ and
$\varphi$ at large energy $s$, or equivalently at $\gamma \gg 1$, as
\be
f = -\gamma \, \Gamma_{\rm cusp}(\as(\lambda^2)) + \CO(\gamma^0)\,,
\qquad
\varphi \sim \Gamma_{\rm cross}^{22} = \CO(\gamma)\,,
\lab{f-as}
\ee
where the explicit expression for the leading term in $\varphi$ will not be
important.

Let us first consider the evolution equation \re{eq1} is the LLA,
$\as\ln s \sim 1$ and $\as \ll 1$. In this limit, one can neglect
the $\lambda-$dependence of $\Gamma_{\rm cross}$ and omit in \re{f}
the terms involving the derivatives of the matrix elements.
The eigenvalues \re{+-} of the matrix $\Gamma_{\rm cross}$ become
$$
\Gamma_+^{\rm LLA}{=}\alpi N_c\ln\frac{s}{m^2}
\,,
\qqquad \Gamma_-^{\rm LLA}{=}0
$$
and using the identities $\det\Gamma_{\rm cross}=\Gamma_+ \Gamma_-$ and
$\tr\Gamma_{\rm cross}=\Gamma_+ + \Gamma_-$ one can obtain the
solution of \re{eq1} in the LLA limit as
\be
T_2^{\rm LLA} 
=
\frac{2\pi\widetilde C_2}{q_\perp^2}\
\lr{\frac{s}{m^2}}^{1-\frac{\as}{2\pi} N_c\ln\frac{q^2}{\lambda^2}}
\lab{LLA}
\ee
with $\widetilde C_2 \equiv \widetilde C_2(\gamma,1)$ being the integration
constant independent of the IR cutoff. Comparing \re{LLA} with \re{T2-},
\re{Born}
and \re{1-loop} we find that the LLA solution of the evolution equation
\re{eq1} correctly reproduces the one-loop correction to the Reggeon
trajectory \re{1-loop} provided that $\widetilde C_2=\as$ in accordance
with \re{C2t}.

Let us now go beyond the LLA order and try to solve the evolution
equation \re{eq1} assuming that the gluon is reggeized and the scattering
amplitude $T_2$ has the Regge behaviour
\be
T_2 \sim \lr{\frac{s}{m^2}}^{1+\omega(q^2)}
\equiv
\e^{\gamma(1+\omega(q^2))}
\lab{ans}
\ee
with $\alpha(t)=1+\omega(t)$ being the Reggeon trajectory. To this end,
we substitute the ansatz \re{ans} into the evolution equation \re{eq1}
and obtain the Ricatti equation for
$\omega'\equiv \frac{d}{d\ln\lambda}\omega$
$$
\frac1{\gamma} \omega'' + \lr{\omega'}^2
+\frac{f}{\gamma}\ \omega' 
+\frac{\varphi}{\gamma^2} 
=0\,.
$$
Using \re{f-as} together with $\omega  = \CO(\gamma^0)$, we notice that
the first and the last terms in this equation are suppressed by
a power of $\gamma$ and the third term is proportional to
$\Gamma_{\rm cusp}$. Therefore in the leading $\gamma\to \infty$ limit
the evolution equation is reduced to
\be
\omega' \equiv \lambda \frac{d \omega}{ d\lambda}  =
\Gamma_{\rm cusp}(\as(\lambda^2))
\lab{eq2}
\ee
with $\omega$ being the function of $q^2/\lambda^2$ and $\as(\lambda^2)$.
Integrating the evolution equation \re{eq2} we obtain the Reggeon
trajectory as
\be
\alpha_\R(t)-1=
\omega(q_\perp^2)
= \Gamma_\R(\as(q_\perp^2)) - \frac12
\int_{\lambda^2}^{q_\perp^2}
\frac{dk_\perp^2}{k_\perp^2} \Gamma_{\rm cusp}(\as(k_\perp^2))\,,
\lab{all}
\ee
or equivalently
\be
\omega(q_\perp^2)=\Gamma_\R(\as(q_\perp^2)) -
\int_{\as(\lambda^2)}^{\as(q_\perp^2)}d\as
\frac{\Gamma_{\rm cusp}(\as)}{\beta(\as)}\,.
\lab{all1}
\ee
Here, the anomalous dimension $\Gamma_\R$ appears as the integration
constant.

Using \re{all} one can expand the Reggeon trajectory as a double series in
$\as(\lambda^2)$ and $L=\frac12\ln\frac{q_\perp^2}{\lambda^2}$. The fact that
$\omega(q_\perp^2)$ satisfies the RG equation \re{eq2} imposes
severe restrictions on this expansion. In particular, the leading
term of the series has a form $(\as L)^n$ with the coefficient
in front proportional to a power of the $\beta-$function. This explains the
origin of the $\as^2 L^2$ term and the cancellation of $\as^2 L^3$
term in \re{2-loop}. Moreover, one
immediately identifies the coefficient $w_2$ in front of the $\as^2 L$
term in \re{2-loop} as the 2-loop correction to the cusp anomalous
dimension \re{cusp}. As to remaining coefficient $w_3$ in \re{2-loop},
it provides the 2-loop correction to the anomalous dimension $\Gamma_\R$
\be
\Gamma_\R= \lr{\alpi}^2 N_c\left[
\lr{\frac{101}{108}-\frac18\zeta(3)}N_c-\frac7{54}n_f
\right] +\CO(\as^3)\,.
\lab{G_R}
\ee
The following concluding remarks are in order.
The solution of the evolution equation, \re{all} and \re{all1},
gives the expression for the Reggeon trajectory in terms of
three functions of the coupling constant: the $\beta-$function,
the cusp anomalous dimension, $\Gamma_{\rm cusp}$, and
the anomalous dimension $\Gamma_\R$.
The universality of the Reggeon trajectory implies that the functions
should not depend on the scattered particles. This is certainly
true for the $\beta-$function but it is not obvious for $\Gamma_{\rm
cusp}$ and $\Gamma_\R$.
Replacing the scattering particles, one has to change correspondingly
the representation of the $SU(N_c)$ group, in which the Wilson lines are
defined in \re{T}. The renormalization properties of the Wilson lines
will also be changed and, in particular, the matrix of the cross
anomalous dimensions will have a different dimension. However,
in the high-energy limit, $\gamma\to \infty$, only the largest eigenvalue
of the matrix will survive. It will coincide with $\Gamma_+$ and will
lead to the evolution equation \re{eq2}.

Since the anomalous dimensions $\Gamma_{\rm cusp}$ and $\Gamma_\R$ 
arose from the analysis of
the quark-quark scattering amplitude, one would naively
expect that they should be defined in the quark representation. However,
their explicit expressions, \re{cusp} and \re{G_R}, do not contain
the corresponding quadratic Casimir operator $C_F$ and, moreover,
$\Gamma_{\rm cusp}$ coincides with the expression \re{cusp} for the cusp
anomalous dimension in the {\it adjoint\/} representation.

The anomalous dimensions $\Gamma_{\rm cusp}$ and $\Gamma_\R$
appeared as peculiar properties of the Wilson lines
describing the high-energy quark--quark scattering. We would like to
notice that the Wilson lines themselves exhibit interesting
universality properties. Calculating perturbative corrections to
Wilson lines describing IR asymptotics of different hard processes
in different kinematics one obtains expressions similar to \re{2-loop}
and \re{all} with the same ``leading'' anomalous dimension
$\Gamma_{\rm cusp}$ and different $\Gamma_\R$. As an example, one
considers the 2-loop calculation of the light-like rectangular
Wilson line $W_\Box$ in the adjoint representation of the $SU(N_c)$
describing the IR asymptotics of the gluon string operator in QCD \ci{KK92}.
Due to additional light-like singularities, the perturbative 
$\CO(\as^2)$ expansion of $\ln W_\Box$ starts with $\as^2 L^3$ term 
with $L$ being the IR logarithm. In a complete analogy with \re{2-loop},
the coefficients in front of the leading and next-to-leading terms
are proportional to the $\beta$ function and $\Gamma_{\rm cusp}$,
respectively, while the next-to-next-to-leading coefficient is
$
\lr{\frac{101}{54}-\frac74\zeta(3)}N_c - \frac 7{27} n_f\,.
$
Remarkably enough, this expression coincides with $2 w_3$ defined in
\re{2-loop} up to $\zeta(3)-$term. We believe that the agreement is
not incidental.

Our calculation of the scattering amplitude was entirely perturbative
and it relied on the fact that the transferred momentum $q_\perp^2$
and the IR cutoff have been  chosen in \re{kin} to be large enough for the
perturbative expansion in \re{all} to be  meaningful. Once $q_\perp^2$
and $\lambda^2$ are decreasing toward $\Lambda_{\sc qcd}^2$ one
should expect the emergence of nonperturbative corrections to the
Reggeon trajectory. Their general structure can be predicted by
analysing ambiguities of the perturbative series related to the
contribution of the IR renormalons \ci{K95,Le,ARS}.
Using the nonperturbative definition \re{T} of the scattering amplitude
as Fourier transformed Wilson line expectation value, one can identify
the leading IR renormalon contribution to the scattering amplitude
and to the Reggeon trajectory \ci{K95}.

The Reggeon trajectory \re{all} provides an additive correction to the
BFKL kernel. In the BFKL Pomeron, built from two Reggeons, IR divergences 
of the Reggeon trajectory, described by the last term in the r.h.s.\ of 
\re{all}, are effectively cancelled with the contribution to the BFKL 
kernel due to Reggeon-Reggeon interaction. It is the finite part
left after the cancellation which provides the correction to the BFKL 
Pomeron. Similar situation occurs in the Drell-Yan process, in which
the Sudakov form factor plays a role of the gluon Regge trajectory.
In this case, large finite perturbative corrections to the cross section 
can be resummed to all orders \ci{SCT} and the analysis \ci{KM} similar 
to the one performed in the previous sections allows to establish their
intrinsic connection to the cusp singularities of the Wilson lines.
This suggests that there should exist a relation between the nonleading 
corrections to the BFKL Pomeron and certain properties of the Wilson
lines (see e.g. \ci{B}).


\bigskip\bigskip
\noindent{\Large{\bf Acknowledgements}}
\bigskip

\noindent
One of us (G.P.K.) is most grateful to V.N. Gribov for stimulating discussions.

\bb{99}
\bi{Gr}    V.N. Gribov, Sov. Phys. JETP 26 (1968) 414; Nucl. Phys. B106
           (1976) 189.
\bi{BFKL}  V.S. Fadin, E.A. Kuraev and L.N. Lipatov, Phys. Lett. B60 (1975) 50;
           Sov. Phys. JETP 44 (1976) 443; 45 (1977) 199;
\\         Ya.Ya. Balitsky and L.N. Lipatov, Sov. J. Nucl. Phys. 28 (1978) 822.
\bi{CSS}   J.C. Collins, D.E. Soper and G. Sterman, in
           {\it Perturbative QCD\/},  ed. A.H. Mueller, World Scientific
           Publ., 1989.
\bi{BS89}  J. Botts and G. Sterman, Nucl. Phys. B325 (1989) 62;
           Phys. Lett. B224 (1989) 201.
\bi{K94}   G.P. Korchemsky, Phys. Lett. B325 (1994) 459.
\bi{SS94}  M.G. Sotiropoulos and G. Sterman, Nucl. Phys. B425 (1994) 489;
           B419 (1994) 59.
\bi{L}     L.N. Lipatov, Nucl. Phys. B309 (1988) 379.
\bi{FL}    V.S. Fadin and L.N. Lipatov, JETP Lett. 49 (1989) 352; Sov.
           J. Nucl. Phys. 50 (1989) 712; preprint DESY-96-020, Feb 1996;
           [hep-ph/9602287].
\bi{FFK95} V.S. Fadin, R. Fiore and M.I. Kotsky, Phys. Lett. B359 (1995) 181;
\\         V.S. Fadin, R. Fiore and A. Quartarolo, Phys. Rev. D53 (1996) 2729.
\bi{FFK96} V.S. Fadin, R. Fiore and M.I. Kotsky, preprint BUDKER-INP/96-35,
           May 1996; [hep-ph/9605357].
\bi{VV}    E. Verlinde and H. Verlinde, preprint PUPT-1319, Feb 1993;
           [hep-th/9302104].
\bi{KR}    G.P. Korchemsky and A.V. Radyushkin, Phys. Lett. B171 (1986) 459;
           B279 (1992) 359.
\bi{ren}   R.A. Brandt, F. Neri and M.-A. Sato, Phys. Rev. D24 (1981) 879.
\bi{KK95}  I.A. Korchemskaya and G.P. Korchemsky, Nucl. Phys. B437 (1995) 127.
\bi{BB}    M. Beneke and V.M. Braun, Nucl. Phys. B454 (1995) 253.
\bi{KK92}  I.A. Korchemskaya and G.P. Korchemsky, Phys. Lett. B287 (1992) 169.
\bi{K95}   I.A. Korchemskaya, preprint ITP-SB-95-19, May 1995;
           [hep-ph/9506402].
\bi{Le}    E.M. Levin, Nucl. Phys. B453 (1995) 303.
\bi{ARS}   K.D. Anderson, D.A. Ross and M.G. Sotiropoulos,
           preprint SHEP-96-06, Feb 1996; [hep-ph/9602275].
\bi{SCT}   G. Sterman, Nucl. Phys. B281 (1987) 310; 
\\         S. Catani and L. Trentadue, Nucl. Phys. B327 (1989) 323; B353
           (1991) 183.
\bi{KM}    G.P. Korchemsky and G. Marchesini, Phys. Lett. B313 (1993) 433.
\bi{B}     I. Balitsky, Nucl. Phys. B463 (1996) 99. 

\eb
\end{document}